\documentclass[twocolumn,aps,prb,superscriptaddress,floatfix,longbibliography]{revtex4-1}

\usepackage[latin9]{inputenc}
\usepackage{amsmath}
\usepackage{amssymb}
\usepackage{graphicx}
\usepackage{upgreek}
\usepackage{dsfont}
\usepackage{xcolor}

\newcommand{\vect}[1]{\boldsymbol{\mathbf{#1}}}      
\newcommand{\mathcomma}{~ ,}       
\newcommand{\mathperiod}{~ .}           
\newcommand{\dosf}{\rho_{\mathrm{F}}}     
\DeclareMathOperator{\sgn}{sgn}
\DeclareMathOperator{\tr}{tr}

\begin{document}

\title{Disorder-promoted $C_4$-symmetric magnetic order in iron-based superconductors}

\author{Mareike Hoyer}
\affiliation{Institut für Theorie der Kondensierten Materie, Karlsruher Institut
für Technologie, D-76131 Karlsruhe, Germany}
\affiliation{Institut für Festkörperphysik, Karlsruher Institut für Technologie,
D-76021 Karlsruhe, Germany}

\author{Rafael M.\ Fernandes}
\affiliation{School of Physics and Astronomy, University of Minnesota, Minneapolis
55455, USA}

\author{Alex Levchenko}
\affiliation{Department of Physics, University of Wisconsin-Madison, Madison,
Wisconsin 53706, USA}

\author{Jörg Schmalian}
\affiliation{Institut für Theorie der Kondensierten Materie, Karlsruher Institut
für Technologie, D-76131 Karlsruhe, Germany}
\affiliation{Institut für Festkörperphysik, Karlsruher Institut für Technologie,
D-76021 Karlsruhe, Germany}

\date{\today}

\begin{abstract}
In most iron-based superconductors, the transition to the magnetically ordered state is closely linked to a lowering of structural symmetry from tetragonal ($C_{4}$) to orthorhombic ($C_{2}$). However, recently, a regime of $C_{4}$-symmetric magnetic order has been reported in certain hole-doped iron-based superconductors. This novel magnetic ground state can be understood as a double-\textbf{Q} spin density wave characterized by two order parameters~$\vect{M}_{1}$ and $\vect{M}_{2}$ related to each of the two $\mathbf{Q}$ vectors. Depending on the relative orientations of the order parameters, either a noncollinear spin-vortex crystal or a nonuniform charge-spin density wave could form. Experimentally, Mössbauer spectroscopy, neutron scattering, and muon spin rotation established the latter as the magnetic configuration of some of these optimally hole-doped iron-based superconductors. Theoretically, low-energy itinerant models do support a transition from single-\textbf{Q} to double-\textbf{Q} magnetic 
order, but with nearly-degenerate spin-vortex crystal and charge-spin density wave states. In fact, extensions of these low-energy models including additional electronic interactions tip the balance in favor of the spin-vortex crystal, in apparent contradiction with the recent experimental findings. In this paper, we revisit the phase diagram of magnetic ground states of low-energy multi-band models in the presence of weak disorder. We show that impurity scattering not only promotes the transition from $C_{2}$ to $C_{4}$-magnetic order, but it also favors the charge-spin density wave over the spin-vortex crystal phase. Additionally, in the single-\textbf{Q} phase, our analysis of the nematic coupling constant in the presence of disorder supports the experimental finding that the splitting between the structural and stripe-magnetic transition is enhanced by disorder. 
\end{abstract}

\maketitle

\section{Introduction}

One of the common features of iron-based superconductors (FeSC) is
the emergence of superconductivity in close proximity to a magnetic
instability.~\cite{Paglione.Greene.2010,ChubukovHirschfeld-review2015}
Even more intriguingly, superconductivity coexists with magnetism
in some of the iron-based compounds.~\cite{PhysRevLett.103.087001,PhysRevB.80.140501}
Thus it is imperative to study the nature of the magnetic order in
the FeSC compounds in order to better understand the superconducting
state in these materials.

Most of the undoped compounds of the FeSC family exhibit magnetic
stripe order with the spins on the iron sites lying in the planes
and being aligned ferromagnetically along one direction, and antiferromagnetically
along the other. In addition to the continuous $O(3)$ spin-rotational
symmetry broken below the magnetic transition temperature~$T_{\mathrm{N}}$,
this stripe-magnetic (SM) state also breaks an additional $\mathbb{Z}_{2}$~Ising-like
symmetry since the ordering vector of the spin-density wave (SDW)
$\vect{S}(\vect{r})=\vect{M}\mathrm{e}^{\mathrm{i}\vect{Q}\cdot\vect{r}}$
is either $\vect{Q}=(0,\pi)$ or $\vect{Q}=(\pi,0)$. The $\mathbb{Z}_{2}$ (or, equivalently, $C_2$)~symmetry breaking can occur at temperatures~$T_{\mathrm{s}}>T_{\mathrm{N}}$
and entails a structural transition from tetragonal ($C_{4}$) to
orthorhombic ($C_{2}$). Furthermore, if the transitions are split,
this allows for an intermediate phase with broken $\mathbb{Z}_{2}$~symmetry
but no magnetic long-range order. This intermediate phase is dubbed
\emph{nematic order}~\cite{ChenKivelson-PRB2008,CenkeMuellerSachdev-PRB2008,Fernandes2014-Review}. Interestingly, the splitting $\Delta T=T_{\mathrm{s}}-T_{\mathrm{N}}$
between the two transitions, and the stabilization of an intermediate
nematic phase, depends on disorder.~\cite{JescheGeibel-PRB2010,Shibauchi-2015,LiangDagotto-PRB2015}.
Uncovering the origin of the nematic phase -- either a spin-driven
or an orbital-driven mechanism -- may also elucidate the mechanism
for superconductivity. 

\begin{figure*}[t]
\includegraphics[width=0.8\textwidth]{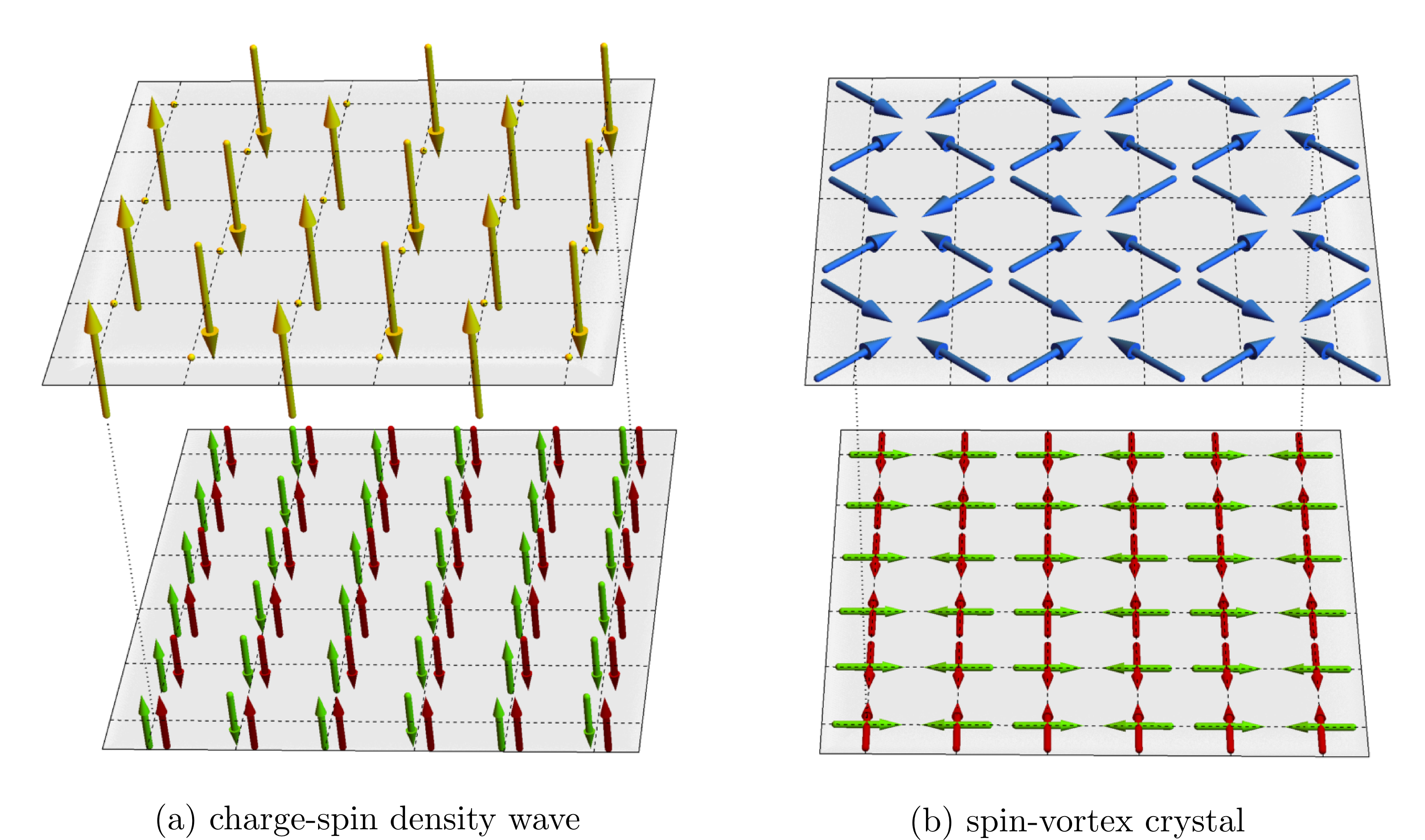}
\caption{Illustration of the two double-$\vect{Q}$ magnetically ordered states
as a superposition of two single-$\vect{Q}$ stripe-magnetic states.
(a) Aligning the order parameters $\vect{M}_{1}=\pm\vect{M}_{2}$
(anti)parallel yields a charge-spin density wave. This order is favorable
if $g<|w|$ and $w<0$. (b) Aligning the order parameters $\vect{M}_{1}\perp\vect{M}_{2}$
perpendicular to each other leads to the formation of a spin-vortex
crystal. This state is favorable if $g<0$ and $w>0$. Otherwise,
single-$\vect{Q}$ stripe order is favored. }
\label{fig:magnetic-ground-states} 
\end{figure*}

Recently, $C_{4}$-magnetic phases have been observed in the hole-doped
compounds Ba(Fe$_{1-x}$Mn$_{x}$)$_{2}$As$_{2}$,~\cite{Kim-PRB2010}
Ba$_{1-x}$Na$_{x}$Fe$_{2}$As$_{2}$,~\cite{Avci-ncomms2014},
Ba$_{1-x}$K$_{x}$Fe$_{2}$As$_{2}$,~\cite{Hassinger-PRB2012,Bohmer-ncomms2015,AllredChmaissem-arxiv2015}
and Sr$_{1-x}$K$_{x}$Fe$_{2}$As$_{2}$ \cite{Allred-arxiv2015},
suggesting that such phases might be a general feature in the phase
diagram of hole-doped FeSC~\cite{Wang-PRB2015}. The magnetic Bragg
peaks of these $C_{4}$-magnetic phases occur at the same momenta
$\vect{Q}_{1}=(\pi,0)$ and $\vect{Q}_{2}=(0,\pi)$ as in the stripe-ordered
state and, consequently, such a state can be understood as the superposition
of two spin-density waves~$\vect{S}(\vect{r})=\vect{M}_{1}\mathrm{e}^{\mathrm{i}\vect{Q}_{1}\cdot\vect{r}}+\vect{M}_{2}\mathrm{e}^{\mathrm{i}\vect{Q}_{2}\cdot\vect{r}}$,
i.\,e. a double-\textbf{Q} SDW, as illustrated in Fig.~\ref{fig:magnetic-ground-states}.
As in the case of stripe antiferromagnetism, which is preceded by
nematic order, also these double-\textbf{Q} magnetic states can in
principle be melted in two stages, passing through an intermediate
state of vestigial charge or chiral order.~\cite{FernandesKivelsonBerg2015}

The existence of double-$\vect{Q}$ magnetic states as additional
ground states for the FeSC has also been established by different
theoretical approaches,~\cite{Lorenzana-PRL2008,EreminChubukov-PRB2010,KangTesanovic-PRB2011,Giovannetti-ncomms2011,Brydon-PRB2011,Fernandes2012,CvetkovicVafek-PRB2013,Kang-PRB2015,GastiasoroAndersen-PRB2015}
all of which suggest the two possible double-$\vect{Q}$ ground states
visualized in Fig.~\ref{fig:magnetic-ground-states} in addition
to the single-$\vect{Q}$ stripe-magnetic order. Fig.~\ref{fig:magnetic-ground-states}(a)
shows the \emph{charge-spin density wave} (CSDW) that arises from
aligning $\vect{M}_{1}$ and $\vect{M}_{2}$ either parallel or antiparallel.
This results in a nonuniform magnetization with vanishing average
moment at the even lattice sites and staggered-like order at the odd
lattice sites, or vice versa. If $\vect{M}_{1}$ and $\vect{M}_{2}$
are orthogonal, the resulting \emph{spin-vortex crystal} (SVC) is
characterized by a noncollinear magnetization that is illustrated
in Fig.~\ref{fig:magnetic-ground-states}(b).

All three magnetic states, the stripe-magnetic and the two double-$\vect{Q}$
magnetic states, can be rationalized in terms of a Ginzburg-Landau
expansion of the free energy in terms of the two magnetic order parameters $\vect{M}_{i}$~\cite{Fernandes2012,WangFernandes-PRB2014,FernandesKivelsonBerg2015}, 
\begin{align}
F[\vect{M}_{i}] & =a\left(\vect{M}_{1}^{2}+\vect{M}_{2}^{2}\right)+\frac{u}{2}\left(\vect{M}_{1}^{2}+\vect{M}_{2}^{2}\right)^{2}\nonumber \\
 & \qquad-\frac{g}{2}\left(\vect{M}_{1}^{2}-\vect{M}_{2}^{2}\right)^{2}+2w\left(\vect{M}_{1}\cdot\vect{M}_{2}\right)^{2}\mathperiod\label{eq:free-energy-from-symmetry}
\end{align}
Depending on the quartic coefficients $u$, $g$, and $w$, the corresponding
energy is minimized by one of the three magnetic ground states described
above, provided that $u>\max(0,g,-w)$.

For $g>\max(0,-w)$, the stripe-ordered $C_{2}$-magnetic phase is
the magnetic ground state of systems described by the free energy~(\eqref{eq:free-energy-from-symmetry}),
and it is accompanied by a structural transition from tetragonal to
orthorhombic. This scenario is supported by itinerant as well as by
localized approaches to magnetism in FeSC, and it is experimentally
well established that stripe-SDW is the magnetic ground state of many
compounds of this family of materials. If, on the other hand, $g<\max(0,-w)$,
one of the two above described possibilities of $C_{4}$-magnetic
phases is realized, depending on whether $w>0$ (leading to a spin
vortex crystal) or $w<0$ (implying a charge-spin density wave).

\begin{figure*}[t]
\includegraphics[width=1\textwidth]{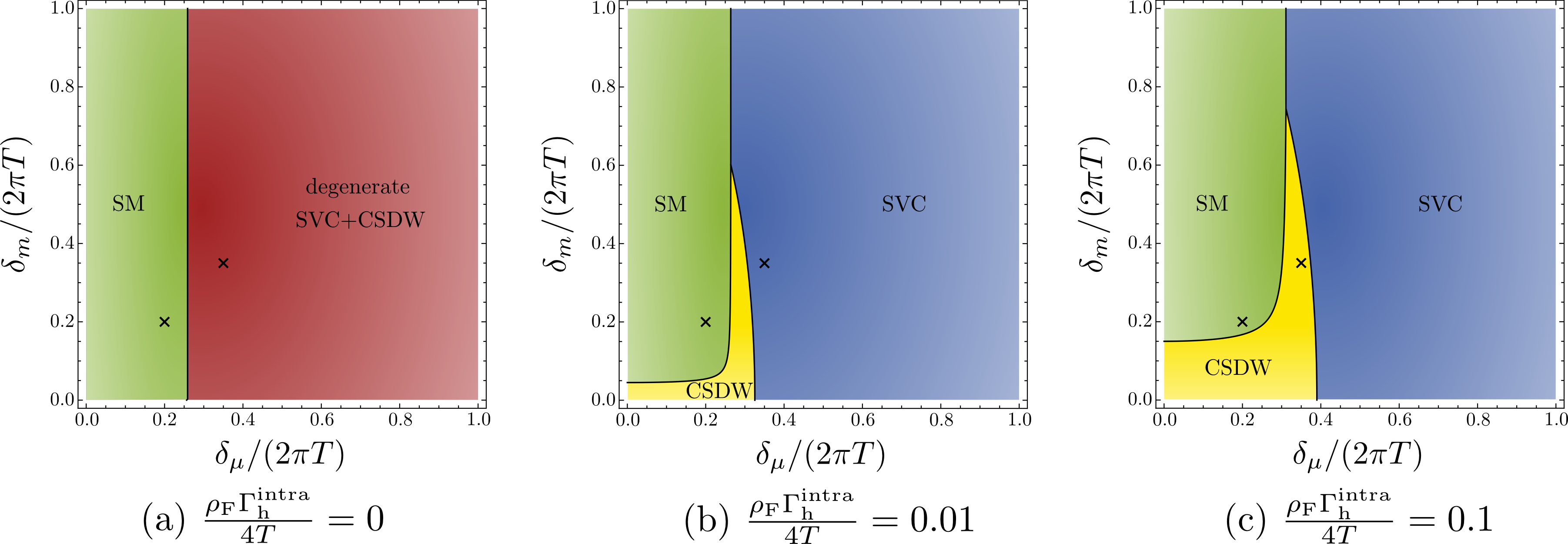}
\caption{Evolution of the phase diagrams, as function of increasing scattering
rate, of the possible magnetic ground states of our three-band minimal
model of iron-based superconductors. Here we used $\Gamma_{\mathrm{e-e}}^{\mathrm{inter}}=0.1\Gamma_{\mathrm{h}}^{\mathrm{intra}}$,
and the phase diagrams are obtained in the limit $\delta_{\mu}\ll2\pi T$
and $\delta_{m}\ll2\pi T$. The regime of single-\textbf{Q} stripe
order (SM) is shown in green, the double-\textbf{Q} spin-vortex crystal
(SVC) order is indicated by blue, and the yellow region represents
the double-\textbf{Q} charge-spin density wave (CSDW). In the clean
regime, where all scattering rates are zero, SVC and CSDW order are
degenerate and we indicated this region with $w=0$ in red. The crosses
mark the points in the phase diagram at which we plotted $g$ and
$w$ as a function of scattering rate in Fig.~\ref{fig:plots-g}
and Fig.~\ref{fig:plots-w}, respectively.}
\label{fig:phasediagrams} 
\end{figure*}

Experimentally, several probes~\cite{Allred-arxiv2015,WasserBraden-PRB2015,Mallett-EPL2015}
established that the magnetic moments in the $C_{4}$-magnetic phase
observed in hole-doped FeSC are aligned parallel to the $c$~axis,
i.\,e., pointing out of plane, and that the magnetic moment vanishes
at every second lattice site while it is doubled at the others. These
features uniquely identify this $C_{4}$-magnetic phase as a realization
of a charge-spin density wave, corresponding to $w<0$. Therefore,
it is important to elucidate theoretically which generic features
of low-energy models yield $w<0$ and $g<|w|$.

Localized approaches based on the $J_{1}$-$J_{2}$~Heisenberg model
favor the single-$\vect{Q}$ stripe-ordered state,~\cite{Chandra-PRL1990}
whereas itinerant approaches allow for both signs of $g$. Focusing
on the three-band itinerant low-energy model previously employed in
the literature \cite{EreminChubukov-PRB2010,Fernandes2012,Wang-PRB2015},
one finds a sign-change from $g>0$ near perfect nesting to $g<0$
away from perfect nesting, implying a transition from single-\textbf{Q}
to a double-$\vect{Q}$ state. However, due to phase space restrictions,
this same model generically gives $w=0$ (for details, see Sect.~\ref{sec:clean-3-band}), leaving the noncollinear
SVC and the nonuniform CSDW order degenerate (see Fig.~\ref{fig:phasediagrams}(a)).
Extensions of this model tend to favor $w>0$, in disagreement with
the recent experiments -- this is indeed obtained by including residual
electronic interactions\cite{EreminChubukov-PRB2010,Wang-PRB2015}
or, as we will show below, an incipient fourth pocket. We note that
although Ref.~\onlinecite{WangFernandes-PRB2014} proposed that the proximity
to a Néel-like state can favor $w<0$, this scenario is only applicable
to Ba(Fe$_{1-x}$Mn$_{x}$)$_{2}$As$_{2}$, since the compound BaMn$_{2}$As$_{2}$
displays Néel order -- which is not the case for Ba$_{1-x}$Na$_{x}$Fe$_{2}$As$_{2}$
or Ba$_{1-x}$K$_{x}$Fe$_{2}$As$_{2}$. Note also that Ref.~\onlinecite{Christensen-arxiv2015}
showed that the spin-orbit coupling leads to anisotropic quadratic
terms in the free energy (\ref{eq:free-energy-from-symmetry}) that
favor the CSDW order, even though $w=0$. This however only works
near the magnetic transition, since at low temperatures the quartic
terms are the ones that determine the ground state.

Therefore, understanding which additional features can lead to $w<0$
is essential to shed light on the mechanisms behind the formation
of the $C_{4}$ phase. Since charged potential impurities can locally
stabilize charge-spin density wave order,~\cite{Lorenzana-PRL2008,GastiasoroAndersen-PRB2015}
one promising approach is the inclusion of doping effects beyond a
rigid-band model. In this paper, we consider
the effect of impurity scattering on the quartic coupling constants
$g$ and $w$ of the itinerant minimal three-band model. We find that,
in the regime where $g>0$ in the clean system, the inclusion of disorder
suppresses $g$, and may even change its sign. One of the consequences
of this result is that the splitting $\Delta T=T_{\mathrm{s}}-T_{\mathrm{N}}$
between the structural and magnetic transitions to the stripe-ordered
state may, depending on the vicinity of the system to a tricritical point, enhance upon increasing disorder, in agreement with recent
experiments on BaFe$_{2}$As$_{2}$ subject to electron irradiation
\cite{Shibauchi-2015}. Furthermore, disorder itself may cause a transition
from single-\textbf{Q} to double-\textbf{Q} order near perfect nesting,
as shown in Figs. \ref{fig:phasediagrams}(b) and (c). Our most important
result, however, is the fate of the vanishing coefficient $w$ in
the presence of disorder. We find that disorder generally lifts the
degeneracy between CSDW and SVC, favoring $w<0$ (and therefore CSDW)
near perfect nesting (Figs. \ref{fig:phasediagrams}(b) and (c)).
Consequently, this opens the interesting possibility of controlling
the magnetic ground state in FeSC with controlled disorder introduced
via irradiation or removed via annealing. 

The paper is structured as follows. Section~\ref{sec:microscopic-model}
introduces the microscopic model including disorder, which we use
to calculate the free energy. We start by recapitulating that $w=0$
follows immediately from a three-band model in Section~\ref{sec:clean-3-band},
followed by the discussion of a fourth band in Section~\ref{sec:clean-4-band}
which only allows for $w>0$. Consequently, we study the effect of
disorder as an alternative route and show in section~\ref{sec:disordered-3-band}
that the presence of impurities can indeed render $w<0$ already within
the simpler three-band model. The nature of the magnetic ground state
is determined by the coefficients $g$ and $w$, and their dependence
on disorder is discussed in Section~\ref{sec:results}, also elucidating
the disorder dependence of the splitting between nematic and magnetic
transition. Finally, we combine our results to obtain a phase diagram
of magnetic ground states in the presence of disorder, which complements
the discussion of our conclusions in Section~\ref{sec:conclusions}.

\section{Microscopic model} 
\label{sec:microscopic-model}

We consider a minimal multi-band model\cite{EreminChubukov-PRB2010,Fernandes2012}
for iron-based superconductors (FeSC) consisting of two circular hole
pockets centered around the $\Gamma$~point and the $M$~point of
the Fe-only Brillouin zone, i.\,e.\ around $(0,0)$ and $(\pi,\pi)$,
respectively, and two elliptical electron pockets centered around
$X$ and $Y$ at $\vect{Q}_{1}=(\pi,0)$ and $\vect{Q}_{2}=(0,\pi)$,
respectively. The pocket at the $M$~point however is not a generic
feature of this class of materials since it exists only in some of
the iron-based compounds. Moreover, even in the compounds in which
the pocket at the $M$~point exists, this band is not guaranteed
to cross the Fermi level for all values of $k_{z}$. Our analysis
in section~\ref{sec:clean-4-band} will show that the presence of
such an incipient hole pocket at the $M$~point cannot explain the
formation of a charge-spin density wave and hence can be neglected
in the remainder of this paper. The noninteracting part of the model
is described by the Hamiltonian 
\begin{align}
\mathcal{H}_{0}=\sum_{\vect{k},\sigma,\lambda}\varepsilon_{\lambda,\vect{k}}c_{\lambda,\vect{k},\sigma}^{\dagger}c_{\lambda,\vect{k},\sigma}^ {}\mathcomma\label{eq:H_0}
\end{align}
where the fact that the bands are centered around different momenta
is reflected in the band index~$\lambda\in\{\mathrm{h}_{\Gamma},\mathrm{h}_{M},\mathrm{e}_{1},\mathrm{e}_{2}\}$
where the hole bands are labeled by $\mathrm{h}_{\Gamma}$ and $\mathrm{h}_{M}$,
and the electron bands by $\mathrm{e}_{1}\equiv\mathrm{e}_{X}$ and
$\mathrm{e}_{2}\equiv\mathrm{e}_{Y}$. Thus $c_{\lambda,\vect{k},\sigma}^{\dagger}$
creates an electron in band~$\lambda$ with spin~$\sigma$, and
the respective dispersions near the Fermi energy can be parametrized
as follows 
\begin{align}
\begin{split} & \varepsilon_{\mathrm{h}_{\Gamma},\vect{k}}=-\varepsilon_{\vect{k}}\mathcomma\\
 & \varepsilon_{\textrm{e}_{1},\vect{k}}=\varepsilon_{\vect{k}}-\delta_{\mu}+\delta_{m}\cos(2\theta)\mathcomma\\
 & \varepsilon_{\textrm{e}_{2},\vect{k}}=\varepsilon_{\vect{k}}-\delta_{\mu}-\delta_{m}\cos(2\theta)\mathcomma\\
 & \varepsilon_{\mathrm{h}_{M},\vect{k}}=-\varepsilon_{\vect{k}}-E_{M}\mathcomma
\end{split}
\end{align}
with $\varepsilon_{\vect{k}}=\frac{k^{2}}{2m}-\varepsilon_{0}+\mu$
and $\theta=\arctan(k_{y}/k_{x})$. $\delta_{\mu}$ characterizes the shift of the chemical potential and is therefore
proportional to doping, and $\delta_{m}$ is a measure of the ellipticity of the electron bands. The top of the hole band at the $M$~point is lower
in energy than the top of the hole band at the $\Gamma$ point, i.e.
$E_{M}>0$, such that it is not guaranteed to cross the Fermi surface
even if it does exist. Note that $E_{M}=\delta_{\mu}=\delta_{m}=0$
gives perfectly nested electron and hole bands. The respective noninteracting
single-particle Green's functions are given by $G_{\lambda,\vect{k}}(\nu_{n})=(\mathrm{i}\nu_{n}-\varepsilon_{\lambda,\vect{k}})^{-1}$
with $\nu_{n}=2\pi T(n+1/2)$ being a fermionic Matsubara frequency.

Since we are concerned with the nature
of the magnetically ordered state, we focus on the electron-electron
interaction projected in the spin channel, hereafter denoted by $V$.
Upon performing a Hubbard-Stratonovich transformation, two magnetic
order parameters arise, $\vect{M}_{1}$ and $\vect{M}_{2}$, associated
with the two ordering vectors $\vect{Q}_{1}=(\pi,0)$ and $\vect{Q}_{2}=(0,\pi)$,
respectively. Their coupling to the electronic degrees of freedom
is given by 
\begin{align}
\mathcal{H}_{\mathrm{int}} & =-\sum_{\vect{k},i}\vect{M}_{i}\cdot\left(c_{\mathrm{h}_{\Gamma},\vect{k},\sigma_{1}}^{\dagger}\vect{\sigma}_{\sigma_{1},\sigma_{2}}c_{\mathrm{e}_{i},\vect{k},\sigma_{2}}^ {}+\mathrm{h.\,c.}\right)\nonumber\\
 & \qquad -\sum_{\vect{k},i}\vect{M}_{i}\cdot\left(c_{\mathrm{h}_{M},\vect{k},\sigma_{1}}^{\dagger}\vect{\sigma}_{\sigma_{1},\sigma_{2}}c_{\mathrm{e}_{\bar{i}},\vect{k},\sigma_{2}}+\mathrm{h.\,c.}\right)
\label{eq_H_int}
\end{align}
where $\bar{i}=2,1$ if $i=1,2$. In the vicinity of the magnetic
phase transition, we can integrate out the electronic degrees of freedom
and derive the free energy expansion of the system 
\begin{align}
F[\vect{M}_{i}] & = \sum_{i}a_{i}|\vect{M}_{i}|^{2}+\sum_{i,j}u_{ij}|\vect{M}_{i}|^{2}|\vect{M}_{j}|^{2} \nonumber\\
 & \qquad+  2w(\vect{M}_{1}\cdot\vect{M}_{2})^{2}\mathcomma\label{eq:free-energy-from-model}
\end{align}
where the coefficients $a_{i}$, $u_{ij}$ and $w$ can be calculated
from the microscopic model introduced above. Due to the rotational
symmetry connecting the electron bands, it holds that $a_{1}=a_{2}$,
$u_{11}=u_{22}$, and $u_{12}=u_{21}$. The free energy~\eqref{eq:free-energy-from-model}
can be brought to the form of Eq.~\eqref{eq:free-energy-from-symmetry}
using $u\equiv u_{12}+u_{11}$ and $g\equiv u_{12}-u_{11}$. 

While the transition temperature is determined by the vanishing of
the quadratic coefficient, the nature of the magnetic ground state
is solely determined by the interplay of the quartic coefficients
$g$ and $w$ in this expansion as long as $u>\max(0,g,-w)$. Since
our goal is to explain the formation of charge-spin density waves
in a low-energy model of the FeSC, we are mainly interested in scenarios
that yield $w<0$. In the remainder of this section, we show that
neglecting the incipient hole pocket at $M$ yields $w=0$ in the
clean case as a consequence of phase space restrictions. However,
including the incipient pocket in a clean model leads to $w>0$ and
thus the spin-vortex crystal would be favorable. Only the inclusion
of disorder can yield $w<0$ and thus render the nonuniform charge-spin
density wave order favorable

\subsection{Clean three-band model}
\label{sec:clean-3-band}

\begin{figure}[b]
\centering 
\includegraphics[width=0.9\columnwidth]{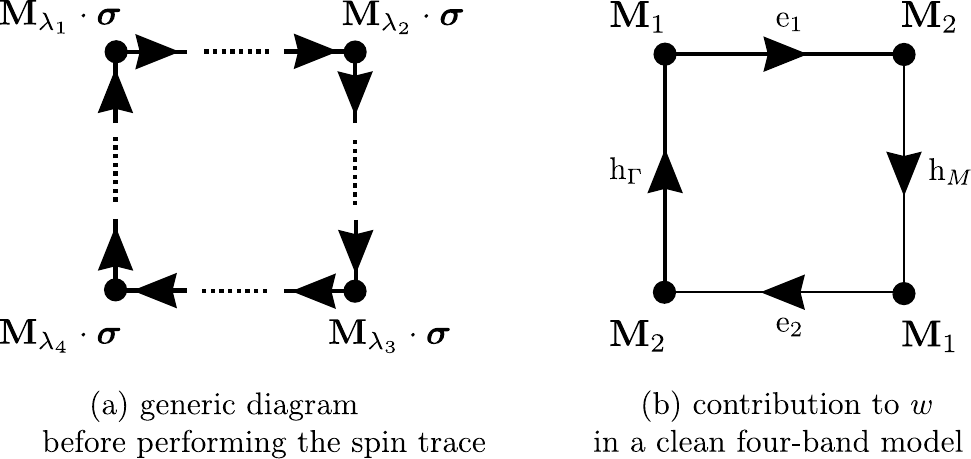}
\caption{Diagrams in the absence of disorder. (a) Sketch of a generic quartic
diagram before performing the spin trace: Each vertex couples the
hole band to one of the electron bands, and the dashed lines indicate
that scattering or additional interactions could alter the diagram.
(b) Contribution to the planar coupling constant $w$ if the incipient
hole pocket at $M$ is taken into account.}
\label{fig:clean-diagrams} 
\end{figure}

We start our considerations with the clean three-band model, i.\,e.,
disregarding the second hole pocket at the $M$~point which is not
present in all FeSC compounds. The coefficients in the expansion of
the free energy, previously defined in Ref.~\onlinecite{Fernandes2012},
are given by: 
\begin{align}
 a_{i} &=\frac{2}{V}+2\int_{k}G_{\mathrm{h}_{\Gamma},\vect{k}}(\nu_{n})G_{\mathrm{e}_{i},\vect{k}}(\nu_{n})\mathcomma \nonumber \\
 u &=\frac{1}{2}\int_k G_{\mathrm{h}_\Gamma,\vect{k}}^2(\nu_n)[G_{\mathrm{e}_1,\vect{k}}(\nu_n)+G_{\mathrm{e}_2,\vect{k}}(\nu_n)]^2\mathcomma \nonumber \\
 g &=-\frac{1}{2}\int_{k}G_{\mathrm{h}_{\Gamma},\vect{k}}^{2}(\nu_{n})[G_{\mathrm{e}_{1},\vect{k}}(\nu_{n})-G_{\mathrm{e}_{2},\vect{k}}(\nu_{n})]^{2}\mathcomma \nonumber \\
 w &=0\mathcomma
\end{align}
where we abbreviated $\int_{k}\ldots\equiv T\sum_{n}\int\frac{\mathrm{d}\vect{k}}{(2\pi)^{2}}\ldots$
For convenience, we write $u$ and $g$ here in the symmetrized form, namely
$u=\tfrac{1}{2}(u_{11}+u_{12}+u_{21}+u_{22})$ and $g=-\tfrac{1}{2}(u_{11}-u_{12}-u_{21}+u_{22})$. The coefficient $w$ vanishes in the
clean model as a consequence of the trace in spin space \cite{Christensen-arxiv2015}:
The most generic quartic diagram [see Fig.~\ref{fig:clean-diagrams}(a)]
is proportional to 
\begin{align}
 & \tr\bigg[\sum_{ijkl}M_{\lambda_{1}}^{(i)}\sigma_{i}M_{\lambda_{2}}^{(j)}\sigma_{j}M_{\lambda_{3}}^{(k)}\sigma_{k}M_{\lambda_{4}}^{(l)}\sigma_{l}\bigg] \nonumber\\
 & =2\big[(\vect{M}_{\lambda_{1}}\cdot\vect{M}_{\lambda_{2}})(\vect{M}_{\lambda_{3}}\cdot\vect{M}_{\lambda_{4}}) \nonumber\\
 & \qquad-(\vect{M}_{\lambda_{1}}\cdot\vect{M}_{\lambda_{3}})(\vect{M}_{\lambda_{2}}\cdot\vect{M}_{\lambda_{4}})\nonumber\\
 & \qquad+(\vect{M}_{\lambda_{1}}\cdot\vect{M}_{\lambda_{4}})(\vect{M}_{\lambda_{2}}\cdot\vect{M}_{\lambda_{3}})\big]\mathperiod\label{eq_trace_fourth}
\end{align}
Within the minimal model, introduced in Eqs.~\eqref{eq:H_0}
and \eqref{eq_H_int}, and with the additional simplification of neglecting
the pocket at the $M$~point, the absence of scattering as well as
interactions between the electron bands require that either $\lambda_{1}=\lambda_{2}$
and $\lambda_{3}=\lambda_{4}$, or $\lambda_{1}=\lambda_{4}$ and
$\lambda_{2}=\lambda_{3}$ holds, as can be seen from Fig.~\ref{fig:clean-diagrams}(a).
Both conditions result in $\tr[(\vect{M}_{\lambda_{1}}\cdot\vect{\sigma})(\vect{M}_{\lambda_{2}}\cdot\vect{\sigma})(\vect{M}_{\lambda_{3}}\cdot\vect{\sigma})(\vect{M}_{\lambda_{4}}\cdot\vect{\sigma})]=2|\vect{M}_{\lambda_{1}}|^{2}|\vect{M}_{\lambda_{3}}|^{2}$
and thus imply $w=0$ in the clean case. On the contrary, the inclusion
of interband scattering or interactions between the two electron pockets
at $\vect{Q}_{1}$ and $\vect{Q}_{2}$ allows for contributions where
$\lambda_{1}=\lambda_{3}$ and $\lambda_{2}=\lambda_{4}$, rendering
$w$ finite since then $\tr[(\vect{M}_{\lambda_{1}}\cdot\vect{\sigma})(\vect{M}_{\lambda_{2}}\cdot\vect{\sigma})(\vect{M}_{\lambda_{3}}\cdot\vect{\sigma})(\vect{M}_{\lambda_{4}}\cdot\vect{\sigma})]=2\left[2(\vect{M}_{\lambda_{1}}\cdot\vect{M}_{\lambda_{2}})-|\vect{M}_{\lambda_{1}}|^{2}|\vect{M}_{\lambda_{2}}|^{2}\right]$.

\subsection{Incipient hole pocket at the $M$ point} 
\label{sec:clean-4-band} 

The inclusion of an incipient hole pocket
at $(\pi,\pi)$ allows for contributions that render $w$ finite in
an analogous manner. The contribution to the planar coupling~$w$
that survives the spin trace as a consequence of the presence of the
second hole pocket is depicted diagrammatically in Fig.~\ref{fig:clean-diagrams}(b).
We consider the simplest case where $\delta_{\mu}=0=\delta_{m}$,
i.\,e., perfect nesting of the hole band at the $\Gamma$~point
and the two electron bands, since this yields a finite value for the
planar coupling, 
\begin{align}
w & =4\int_{k}G_{\mathrm{h}_{\Gamma},\vect{k}}(\nu_{n})G_{\mathrm{e}_{1},\vect{k}}(\nu_{n})G_{\mathrm{h}_{M},\vect{k}}(\nu_{n})G_{\mathrm{e}_{2},\vect{k}}(\nu_{n})\nonumber \\
 & =4T\sum_{n=-\infty}^{\infty}\dosf\int_{-\infty}^{\infty}\mathrm{d}\epsilon\,\frac{1}{\mathrm{i}\nu_{n}+\epsilon}\frac{1}{\mathrm{i}\nu_{n}+\epsilon+E_{M}}\frac{1}{(\mathrm{i}\nu_{n}-\epsilon)^{2}}\nonumber \\
 & \approx\left\{ \begin{matrix}\frac{7\dosf\zeta(3)}{2\pi^{2}T^{2}}\approx0.43\frac{\dosf}{T^{2}} & \mathcomma\qquad E_{M}\ll T\mathcomma\\
\frac{4\dosf}{E_{M}^{2}} & \mathcomma\qquad E_{M}\gg T\mathcomma
\end{matrix}\right.
\end{align}
where $\zeta(z)$ is the Riemann zeta~function, and we assumed the
density of states at the Fermi level to be given by a constant $\dosf$
in all bands.

Hence we find that the inclusion of the second hole pocket indeed
lifts the degeneracy of the two double-\textbf{Q} magnetically ordered
states. However, it can only account for the formation of a spin-vortex
crystal since $w>0$ always. Furthermore, if the pocket at $M$ is
shifted to energies far below the Fermi level, we reproduce the results
of the previously discussed three-band model since the coefficient
$w$ vanishes in the limit $E_{M}\rightarrow\infty$, which is the
relevant limit for many of the FeSC compounds.

A positive planar coupling $w>0$ has also been obtained in previous
studies of other extensions of the clean three-band model such as
the perturbative inclusion of additional interactions.~\cite{EreminChubukov-PRB2010,Wang-PRB2015}
This suggests a different route to $w<0$ is needed in order to explain
the formation of the collinear CSDW state within this low-energy model.
In the remainder of this paper, we investigate the effect of disorder
on the magnetic ground state. Furthermore, we neglect the hole pocket
at the $M$~point since it is not a generic feature of the FeSC family
and its inclusion is not able to explain why the nonuniform CSDW is
favored over the noncollinear SVC in the hole-doped compounds.

\subsection{Impurity scattering}
\label{sec:disordered-3-band} 

Impurity scattering will affect both
the magnetic transition temperature, determined by the vanishing of
$a_{i}$, and the nature of the magnetic ground state, determined
by $g$ and $w$. Hereafter, we focus on the latter effect -- the former
gives rise to a suppression of the magnetic transition temperature
with disorder, as shown elsewhere \cite{FernandesChubukov2012,LiangDagotto-PRB2015}.
In the particle-hole symmetric case (perfect nesting), where $\delta_{\mu}=\delta_{m}=0$,
the nematic coupling constant $g$ vanishes. Finite ellipticity $\delta_{m}\neq0$, however,
causes $g$ to be finite. The effect of doping can then partially
be accounted for by a finite value of the chemical potential, i.\,e.,
$\delta_{\mu}\neq0$. 

Meanwhile, doping also introduces disorder, which has a different
effect on the electronic structure than the rigid band shift assumed
by changing the chemical potential.\cite{WadatiElfimovSawatzky-PRL2010} For instance, it has been shown that impurity scattering
can locally stabilize charge-spin density wave order,\cite{GastiasoroAndersen-PRB2015}
thus suggesting that the inclusion of disorder for the itinerant electrons
participating in the formation of the magnetically ordered state is an important
ingredient for the investigation of the CSDW state. Hence we consider
an arbitrary realization of nonmagnetic impurities, thus adding the
term 
\begin{equation}
\mathcal{H}_{\mathrm{dis}}=\sum_{\lambda,\lambda^{\prime}}\sum_{\vect{k},\vect{k}^{\prime}}\sum_{\sigma}\, c_{\lambda,\vect{k},\sigma}^{\dagger}W_{\lambda\lambda^{\prime}}(\vect{k},\vect{k}^{\prime})c_{\lambda^{\prime},\vect{k}^{\prime},\sigma}^ {}
\end{equation}
to the Hamiltonian. As usual, we are not interested in quantities
that depend on the microscopic disorder realization, but rather in
self-averaged physical observables. Therefore, we are interested in
disorder-averaged quantities where all information about the disorder
is encoded in the correlation function 
\begin{align}
 & \left<W_{\lambda_{1}\lambda_{1}^{\prime}}(\vect{k}_{1},\vect{k}_{1}^{\prime})W_{\lambda_{2}\lambda_{2}^{\prime}}(\vect{k}_{2},\vect{k}_{2}^{\prime})\right>_{\mathrm{dis}}\\
 & =\Gamma_{\lambda_{1}\lambda_{1}^{\prime}\lambda_{2}\lambda_{2}^{\prime}}(\vect{k}_{1},\vect{k}_{1}^{\prime},\vect{k}_{2},\vect{k}_{2}^{\prime})\updelta(\vect{k}_{1}+\vect{k}_{2}-\vect{k}_{1}^{\prime}-\vect{k}_{2}^{\prime}+\vect{K})\nonumber 
\end{align}
where $\left<\ldots\right>_{\mathrm{dis}}$ denotes the average over
disorder configurations which restores translation invariance, and
$\vect{K}$ is a vector from the reciprocal lattice. Thus the correlator
constitutes a measure of impurity strength and is proportional to
the scattering rate~$\Gamma$ characterizing the respective scattering
process. These scattering rates depend on the impurity concentration
as well as on the strength of the disorder potential itself.

In the remainder, we concentrate on the simplest type of impurities
and thus assume the disorder to be spatially local, $\updelta$-correlated,
and sufficiently smooth such that the momentum dependence of the scattering
rates can be neglected for momenta from the same pocket of the Fermi
surface. Then, scattering within one band or between two bands is
characterized by constant scattering rates $\Gamma_{\mathrm{e}}^{\mathrm{intra}}$,
$\Gamma_{\mathrm{h}}^{\mathrm{intra}}$ or $\Gamma_{\mathrm{e-e}}^{\mathrm{inter}}$,
$\Gamma_{\mathrm{e-h}}^{\mathrm{inter}}$, respectively. Here we assume
that both electron bands are affected in the same way by impurities,
and thus the respective scattering rates are equal -- consistent with
the tetragonal symmetry of the system. Note that in a multiband model,
the effect of impurity scattering can have subtle consequences\cite{Hoyer-PRB2015}
which we avoid here by requiring that all scattering processes be
characterized by real numbers, i.\,e., the impurities do not break
time-reversal invariance locally. Furthermore, we assume the impurity
potential to be sufficiently weak such that single-particle interference
effects can be neglected. In this case, calculating the self-energy
within the Born approximation is appropriate, resulting in 
\begin{equation}
G_{\lambda,\vect{k}}(\nu_{n})=\left(\mathrm{i}\nu_{n}-\varepsilon_{\lambda,\vect{k}}+\tfrac{\mathrm{i}}{2\tau_{\lambda}}\sgn(\nu_{n})\right)^{-1}\label{G-dis}
\end{equation}
for the propagator in band~$\lambda$ in the presence of impurities.
Here, we introduced the elastic scattering time 
\begin{equation}
\tau_{\lambda}=(2\pi\dosf\Gamma_{\mathrm{total}})^{-1},
\end{equation}
which is determined by the total scattering rate including all intraband
and interband scattering processes that affect propagation of electrons
in band $\lambda$.

\section{Results}
\label{sec:results} 

In multiband systems, the interplay of a multitude
of different intraband and interband scattering processes can affect
physical properties. Fortunately, in the iron-based systems, experiments
as well as ab-initio calculations reveal that not all of them are
equally important.~\cite{Karkin-PRB2009,Li-PRB2011,KirshenbaumPaglione-PRB2012,PhysRevLett.103.057001,Fang2009,Science.336.563,PhysRevLett.108.207003,Herbig2015}
This allows us to devise models of impurity scattering that concentrate
on the dominant scattering processes relevant for the calculation
of $w$ and $g$. Such a simplification allows one to draw conclusions
about the dominant effects that are to be expected due to impurity
scattering, but of course restricts exact quantitative predictions.

For many aspects it is sufficient to discriminate between intraband
and interband scattering processes, and thus it is important to note
that interband scattering (which for example causes pair breaking
in the superconducting state) is much weaker than the dominant intraband
scattering process affecting transport properties.\cite{Karkin-PRB2009,Li-PRB2011,KirshenbaumPaglione-PRB2012} Furthermore,
as demonstrated by transport measurements,\cite{PhysRevLett.103.057001,Fang2009} scanning tunneling microscopy,\cite{Science.336.563} and first-principles
density functional theory calculations,\cite{PhysRevLett.108.207003,Herbig2015} the intraband scattering rate in the hole band exceeds the intraband
scattering rate in the electron bands. For these reasons, we consider 
the following hierarchy of scattering rates in the remainder of the
paper:  
\begin{equation}
\Gamma_{\mathrm{e-e}}^{\mathrm{inter}},\Gamma_{\mathrm{e-h}}^{\mathrm{inter}},\Gamma_{\mathrm{e}}^{\mathrm{intra}}\ll\Gamma_{\mathrm{h}}^{\mathrm{intra}}\mathperiod
\label{eq:inter-ll-intra}
\end{equation}

The main advantage of the minimal three-band model of Section \ref{sec:clean-3-band}
is that it allows for a well-defined perturbative expansion near the perfect-nesting limit ($\delta_{\mu}=\delta_{m}=0$) and the clean limit
($\Gamma_{i}=0$), since in this case $g=w=0$, and the degeneracy
of the magnetic ground state is maximal (i.e. the stripe-magnetic, CSDW, and
SVC phases are all degenerate). Therefore, one can assess qualitatively
how different types of perturbations favor distinct ground states.

\subsection{Effect of disorder on the nematic coupling $g$}

\begin{figure}[b]
\includegraphics[width=1\columnwidth]{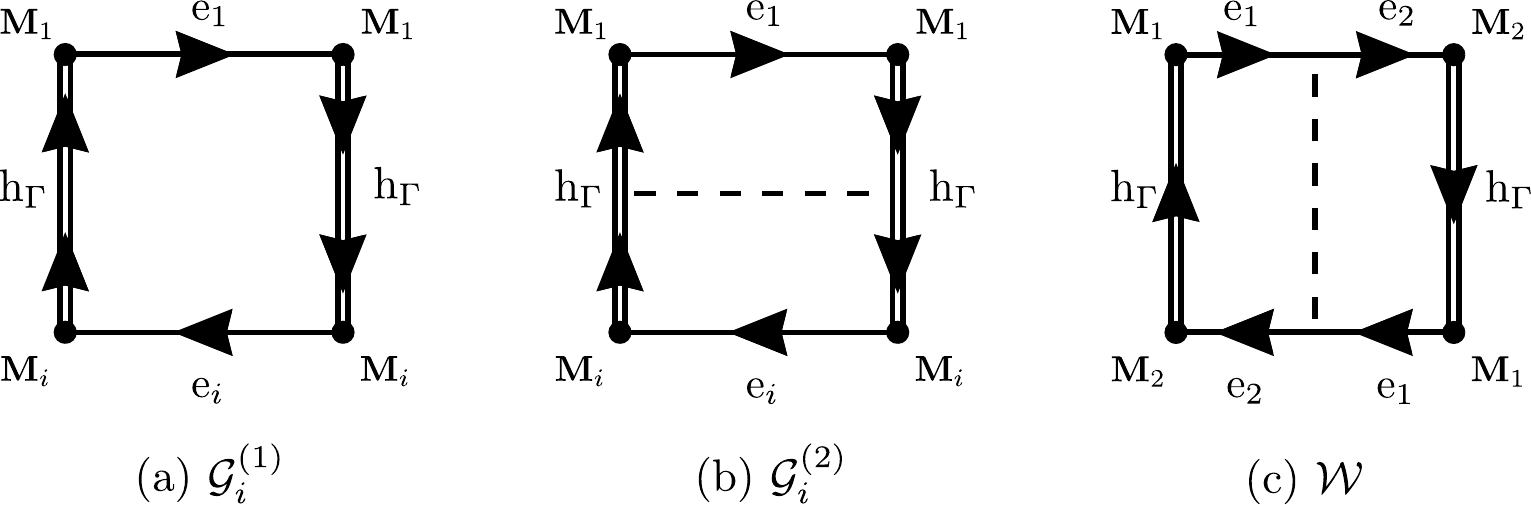} 
\caption{Leading-order diagrams contributing to the quartic coefficients that
determine the magnetic ground state. Double lines indicate that the
respective propagators acquire a finite lifetime due to impurity scattering
whereas single lines are used for propagators in bands that, within
our model, are not affected by impurity scattering. Additional scattering
processes are indicated by a dashed line. (a) and (b) $\mathcal{G}_{i}^{(1)}$ and $\mathcal{G}_{i}^{(2)}$
($i\in\{1,2\}$) are the contributions to $g$ (as well as to $u$)
in the presence of intraband scattering in the hole band which is
the dominant scattering mechanism in FeSC. (c) $\mathcal{W}$ is the
contribution to $w$ which is finite owing to interband scattering
between the two electron bands, and in the presence of the dominant intraband scattering
in the hole band. }
\label{fig:diagrams} 
\end{figure}

We first analyze how disorder affects $g$, since this coupling constant
determines whether the system condenses in a single-\textbf{Q} or
double-\textbf{Q} state. While $g=0$ at perfect nesting, the nematic
coupling constant takes a finite value within the three-band model
as a consequence of the ellipticity of the electron bands; although
orbital dressing effects can make it nonzero even at perfect nesting
\cite{FanfarilloCortijoValenzuela-PRB2015}. Focusing on the contribution
from the dominant scattering rate $\Gamma_{\mathrm{h}}^{\mathrm{intra}}$,
see Eq.~\eqref{eq:inter-ll-intra}, and expanding near perfect nesting,
$\delta_{\mu},\delta_{m}\ll T$, we find  
\begin{widetext}
\begin{align}
 & g=-\frac{T}{2}\sum_{n}\int\frac{\mathrm{d}\vect{k}}{(2\pi)^{2}}G_{\mathrm{h}_{\Gamma},\vect{k}}^{2}(\nu_{n})\left[G_{\mathrm{e}_{1},\vect{k}}(\nu_{n})-G_{\mathrm{e}_{2},\vect{k}}(\nu_{n})\right]^{2}-\Gamma_{\mathrm{h}}^{\mathrm{intra}}\frac{T}{2}\sum_{n}\left[\int\frac{\mathrm{d}\vect{k}}{(2\pi)^{2}}G_{\mathrm{h}_{\Gamma},\vect{k}}^{2}(\nu_{n})[G_{\mathrm{e}_{1},\vect{k}}(\nu_{n})-G_{\mathrm{e}_{2},\vect{k}}(\nu_{n})]\right]^{2}\nonumber \\
 & =\mathcal{G}_{2}^{(1)}-\mathcal{G}_{1}^{(1)}+\mathcal{G}_{2}^{(2)}-\mathcal{G}_{1}^{(2)}
=-\frac{\dosf\delta_{m}^{2}}{1536\pi^{4}T^{4}}\left[\operatorname{\psi_{4}}\left(\frac{1}{2}+\frac{\dosf\Gamma_{\mathrm{h}}^{\mathrm{intra}}}{4T}\right)-\frac{\delta_{\mu}^{2}}{32\pi^{2}T^{2}}\operatorname{\psi_{6}}\left(\frac{1}{2}+\frac{\dosf\Gamma_{\mathrm{h}}^{\mathrm{intra}}}{4T}\right)\right],
\end{align}\end{widetext}
where $\psi_{n}(z)$ is the $n^\textrm{th}$ derivative of the digamma function.
The contributing diagrams $\mathcal{G}_{i}^{(1)}$ and $\mathcal{G}_{i}^{(2)}$
are depicted in Fig.~\ref{fig:diagrams}(a) and (b), and correspond
respectively to the disorder-induced Green's function renormalization
and to the vertex correction. Here, we used that $u_{11}=u_{22}$
and $u_{12}=u_{21}$ holds for the quartic coefficients in the expansion~\eqref{eq:free-energy-from-model}
also in the presence of disorder, and that $\mathcal{G}_{2}^{(2)}-\mathcal{G}_{1}^{(2)}\propto\int\tfrac{\mathrm{d}\theta}{2\pi}\,\cos(2\theta)=0$.

\begin{figure}[t]
\includegraphics[width=0.9\columnwidth]{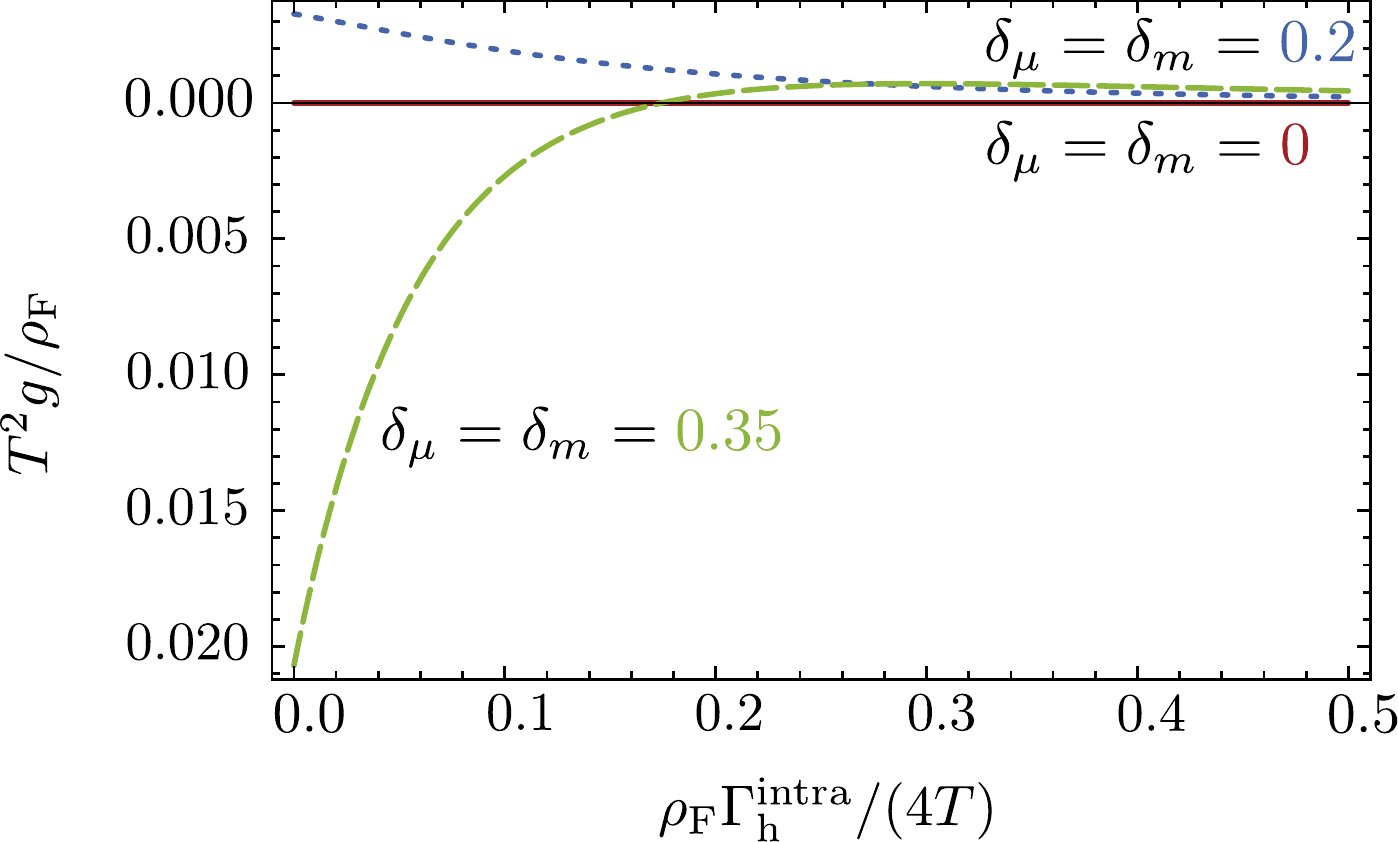}  
\caption{Nematic coupling constant~$g$ in the presence intraband scattering in the
hole band, characterized by the scattering rate $\Gamma_{\mathrm{h}}^{\mathrm{intra}}$. 
We chose $\delta_{\mu}=\delta_{m}=0.2$ (blue, dotted line) as an
example of small ellipticity and detuning which guarantees $w<0$
and $g>0$, and $\delta_{\mu}=\delta_{m}=0.35$ (green, dashed line)
as an example where disorder can tune $g$ and $w$ to be either positive
or negative. The red lines represent the results at particle-hole
symmetry, $\delta_{\mu}=\delta_{m}=0$.}
\label{fig:plots-g} 
\end{figure}

In the clean limit, $\Gamma_{\mathrm{h}}^{\mathrm{intra}}=0$, $g\propto\delta_{m}^{2}$
changes sign from positive to negative for sufficiently large $\delta_{\mu}$,
as shown in Fig.~\ref{fig:phasediagrams}(a) and in agreement
with previous results \cite{Wang-PRB2015}. This describes the transition
from a single-\textbf{Q }to a double-\textbf{Q }magnetic ground state
as the carrier concentration increases. The resulting coupling constant
$g$ as a function of the scattering rate $\Gamma_{\mathrm{h}}^{\mathrm{intra}}$,
plotted for different values of detuning $\delta_{\mu}$ and ellipticity
$\delta_{m}$, is shown in Fig.~\ref{fig:plots-g}. In the particle-hole
symmetric case, $g=0$ as a consequence of $\delta_{m}=0$, regardless
of whether the system is in the clean or dirty limit. Interestingly,
if $g$ is positive (negative) in the clean limit, the addition of
disorder suppresses $g$ and can even induce a sign-change. Therefore,
the transition from a single-\textbf{Q} to a double-\textbf{Q} state
can be controlled not only by carrier concentration, but also by the
disorder potential.

\begin{figure}[t]
\includegraphics[width=0.9\columnwidth]{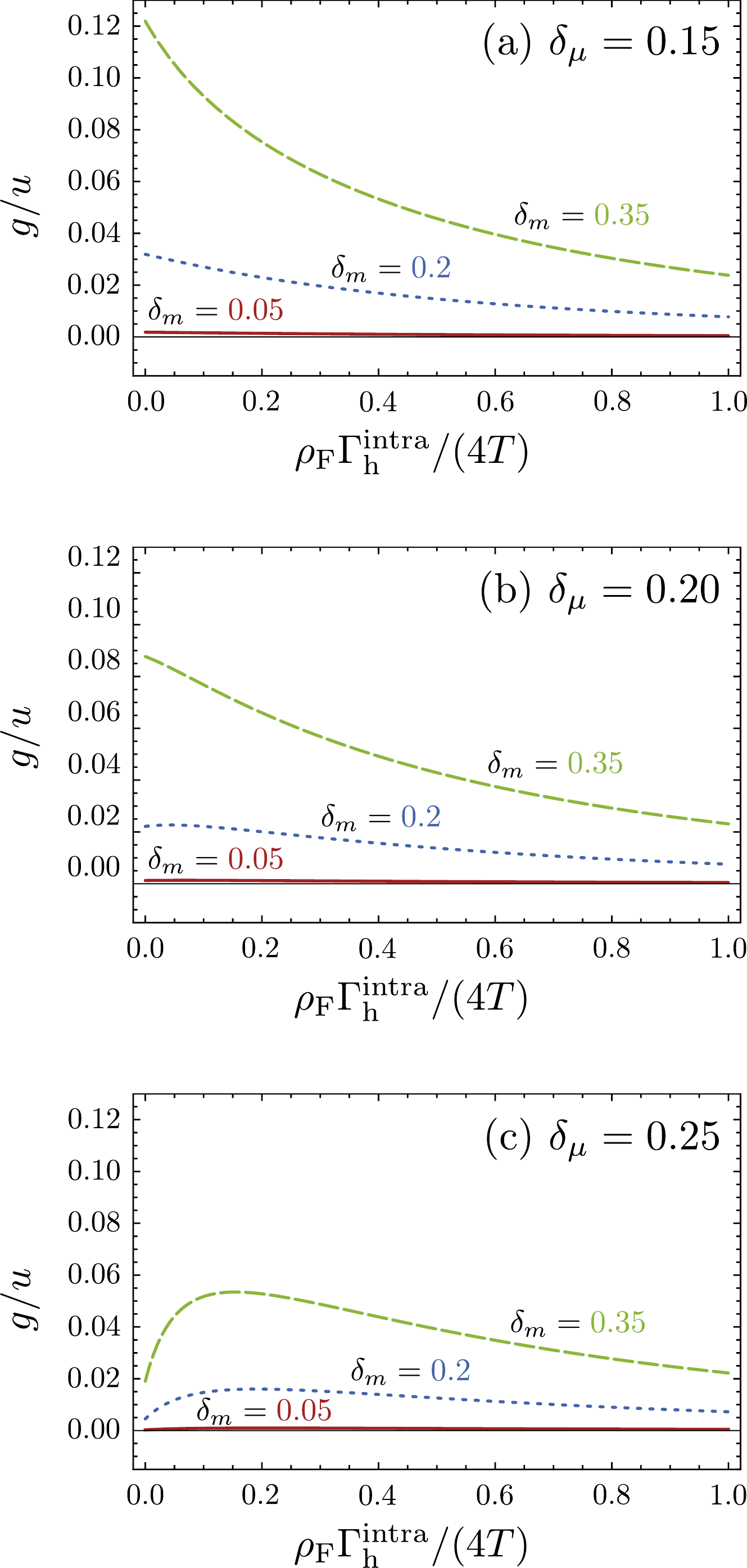} 
\caption{Dependence of the dimensionless nematic coupling constant $g/u$ on
disorder. (a) Close to particle-hole symmetry, $g/u$ decreases monotonically
with increasing scattering rate. (b) and (c) With increasing distance
to particle-hole symmetry, an initial increase of the dimensionless
nematic coupling constant is found for small scattering rates, but
for stronger disorder, the ratio $g/u$ decreases again.}
\label{fig:dimensionless-nematic-coupling} 
\end{figure}

Even when the suppression of $g$ by disorder does not induce a sign-change,
it has important consequences for the phase diagram. In particular,
as shown in Ref.~\onlinecite{Fernandes2012}, the splitting $\Delta T=T_{\mathrm{s}}-T_{\mathrm{N}}$
between the nematic/structural and the magnetic transitions is controlled
by the inverse dimensionless nematic coupling constant $u/g$ and
the dimensionality $d$. In particular, for $2<d<3$, which mimics
an anisotropic 3D system, the two transitions are simultaneous and
first order for $\left(u/g\right)<\left(u/g\right)_{c_{1}}=1/\left(3-d\right)$. For $(u/g)_{c_1}<(u/g)<(u/g)_{c_2}$, the transitions are split and one of them remains first-order whereas the other transition is second-order. In this regime, an increase in $u/g$ results in an enhanced splitting
$\Delta T$, whereas deep in the regime of two split second-order phase
transitions, $\left(u/g\right)\gg\left(u/g\right)_{c_{2}}=\left(6-d\right)/\left(6-2d\right)$,
increasing the ratio $u/g$ reduces the splitting $\Delta T$. To
compute the dimensionless parameter $u/g$, we compute $u$ analogously
to the case of $g$
\begin{widetext}
\begin{align}
  u&=\frac{T}{2}\sum_{n}\int\frac{\mathrm{d}\vect{k}}{(2\pi)^{2}}\, G_{\mathrm{h}_{\Gamma},\vect{k}}^{2}(\nu_{n})\left[G_{\mathrm{e}_{1},\vect{k}}(\nu_{n})+G_{\mathrm{e}_{2},\vect{k}}(\nu_{n})\right]^{2}+\Gamma_{\mathrm{h}}^{\mathrm{intra}}\frac{T}{2}\sum_{n}\left[\int\frac{\mathrm{d}\vect{k}}{(2\pi)^{2}}\, G_{\mathrm{h}_{\Gamma},\vect{k}}^{2}(\nu_{n})[G_{\mathrm{e}_{1},\vect{k}}(\nu_{n})+G_{\mathrm{e}_{2},\vect{k}}(\nu_{n})]\right]^{2}\nonumber \\
 & =\mathcal{G}_{1}^{(1)}+\mathcal{G}_{2}^{(1)}+\mathcal{G}_{1}^{(2)}+\mathcal{G}_{2}^{(2)}
=-\frac{\dosf}{8\pi^{2}T^{2}}\left[\operatorname{\psi_{2}}\Big(\frac{1}{2}+\frac{\dosf\Gamma_{\mathrm{h}}^{\mathrm{intra}}}{4T}\Big)+\frac{\dosf\Gamma_{\mathrm{h}}^{\mathrm{intra}}}{12T}\operatorname{\psi_{3}}\Big(\frac{1}{2}+\frac{\dosf\Gamma_{\mathrm{h}}^{\mathrm{intra}}}{4T}\Big)\right]\nonumber \\
 &\qquad \qquad +\frac{\dosf}{768\pi^{4}T^{4}}\operatorname{\psi_{4}}\Big(\frac{1}{2}+\frac{\dosf\Gamma}{4T}\Big)\left[3\delta_{\mu}^{2}+\delta_{m}^{2}\right]+\frac{\dosf^{2}\Gamma_{\mathrm{h}}^{\mathrm{intra}}}{30720\pi^{4}T^{5}}\operatorname{\psi_{5}}\Big(\frac{1}{2}+\frac{\dosf\Gamma}{4T}\Big)\left[10\delta_{\mu}^{2}+3\delta_{m}^{2}\right]
\end{align}
\end{widetext}
in accordance with previous work.~\cite{Hoyer2014} Near particle-hole
symmetry, where $\delta_{\mu}/(2\pi T)$ and $\delta_{m}/(2\pi T)$
are sufficiently small, and the magnetic ground state is the stripe
one, $g/u$ decreases monotonically with increasing scattering rate
as shown in Fig.~\ref{fig:dimensionless-nematic-coupling}(a). Thus,
if the system initially is near the regime of first-order simultaneous
transitions, as it is the case in undoped BaFe$_{2}$As$_{2}$, the
addition of disorder is expected to cause (or enhance) a splitting
in the magnetic and structural transitions. This agrees with recent
experiments in BaFe$_{2}$As$_{2}$, which observed enhanced splitting
of the transitions upon electron irradiation.\cite{Shibauchi-2015}
This result is also consistent with the theoretical finding of Ref.~\onlinecite{LiangDagotto-PRB2015} that disorder stabilizes the nematic
phase. We note, however, that the dependence of the ratio $g/u$ on
disorder is nonuniversal (see Fig.~\ref{fig:dimensionless-nematic-coupling}(b)
and (c)). In particular, farther away from particle-hole symmetry,
the dependence of $g/u$ on disorder is no longer monotonic: $g/u$
first increases with increasing scattering rate, and above a critical
value starts decreasing again.

\subsection{Effect of disorder on the planar coupling $w$}

Having established that $g$ can become either positive or negative
in both clean and dirty systems, we now analyze $w$. As discussed
above and illustrated in Fig.~\ref{fig:clean-diagrams}(a), in the clean
three-band model $w=0$ always. Following the analysis of the generic
fourth-order diagram in Fig.~\ref{fig:clean-diagrams} and Eq.~\eqref{eq_trace_fourth},
the only scattering processes that gives rise to a nonzero contribution
to $w$ is the one coupling the electron pocket at $\vect{Q}_{1}$
and the electron pocket at $\vect{Q}_{2}$, characterized by the scattering
rate $\Gamma_{\mathrm{e-e}}^{\mathrm{inter}}$. For the sake of clarity,
we neglect all other interband scattering processes since they give
subleading contributions to $w$, i.e. $w=0$ always as long as $\Gamma_{\mathrm{e-e}}^{\mathrm{inter}}=0$.
Then, in the presence of the dominant scattering process, intraband scattering in the hole band and, additionally,
interband scattering between the electron bands, we find  
\begin{align}
 w&=2\Gamma_{\mathrm{e-e}}^{\mathrm{inter}}T\sum_{n}\!\Big[\!\int\!\!\frac{\mathrm{d}\vect{k}}{(2\pi)^{2}}\, G_{\mathrm{h}_{\Gamma},\vect{k}}(\nu_{n})G_{\mathrm{e}_{1},\vect{k}}(\nu_{n})G_{\mathrm{e}_{2},\vect{k}}(\nu_{n})\Big]^{2} \nonumber\\
 & =2\mathcal{W} 
=-\frac{\dosf^{2}\Gamma_{\mathrm{e-e}}^{\mathrm{inter}}}{96\pi^{2}T^{3}}\left[\operatorname{\psi_{3}}\Big(\frac{1}{2}+\frac{\dosf(\Gamma_{\mathrm{h}}^{\mathrm{intra}}+\Gamma_{\mathrm{e-e}}^{\mathrm{inter}})}{4T}\Big)\right. \nonumber \\
 &\qquad \qquad \left.-\frac{10\delta_{\mu}^{2}+\delta_{m}^{2}}{320\pi^{2}T^{2}}\operatorname{\psi_{5}}\Big(\frac{1}{2}+\frac{\dosf(\Gamma_{\mathrm{h}}^{\mathrm{intra}}+\Gamma_{\mathrm{e-e}}^{\mathrm{inter}})}{4T}\Big)\right]\mathcomma
\end{align}
where we assumed the density of states at the Fermi
surface to be given by a constant $\dosf$ in all three bands, and
we expanded to leading order in $\delta_{\mu}$ and $\delta_{m}$
to obtain the results. The respective diagram denoted by $\mathcal{W}$
is depicted in Fig.~\ref{fig:diagrams}(c). Note that contributions
with more than one scattering process between electron bands vanish
upon momentum integration and thus the above result already includes
contributions up to infinite order in $\Gamma_{\mathrm{e-e}}^{\mathrm{inter}}$. 

\begin{figure}[t]
\includegraphics[width=0.9\columnwidth]{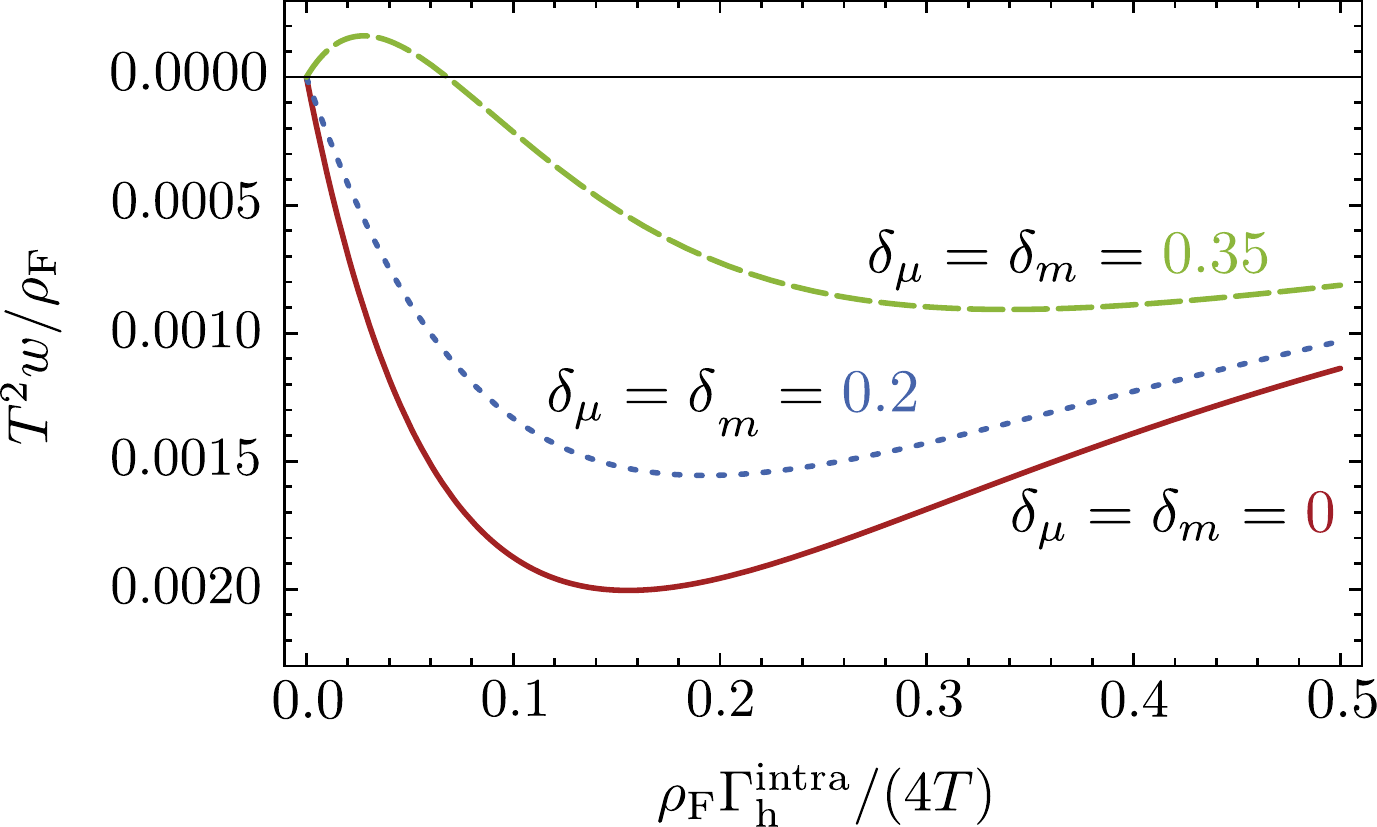}  
\caption{Planar coupling~$w$ as a function of intraband scattering rate in the
hole band, $\Gamma_{\mathrm{h}}^{\mathrm{intra}}$, where we set $\Gamma_{\mathrm{e-e}}^{\mathrm{inter}}=0.1\Gamma_{\mathrm{h}}^{\mathrm{intra}}$.
We chose $\delta_{\mu}=\delta_{m}=0.2$ (blue, dotted line) as an
example of small ellipticity and detuning which guarantees $w<0$
and $g>0$, and $\delta_{\mu}=\delta_{m}=0.35$ (green, dashed line)
as an example where disorder can tune $w$ and $g$ to be either positive
or negative. The red lines represent the results at particle-hole
symmetry, $\delta_{\mu}=\delta_{m}=0$.}
\label{fig:plots-w} 
\end{figure}

We show the coefficient $w$ as a function of the scattering rate
for different values of detuning~$\delta_{\mu}$ and ellipticity~$\delta_{m}$
in Fig.~\ref{fig:plots-w}. In the absence of impurity scattering,
we recover $w=0$. At particle-hole symmetry, $\delta_{\mu}=\delta_{m}=0$,
disorder leads to $w<0$, thus favoring the formation of a charge-spin
density wave (see Fig.~\ref{fig:magnetic-ground-states}(a)) as long
as $g<|w|$. In contrast, finite detuning and ellipticity yield a
contribution of opposite sign and thus, depending on the scattering
rate and the distance from particle-hole symmetry, $w$ can be either
positive or negative, allowing for both proposed double-$\vect{Q}$
states, CSDW and the SVC. This conclusion holds also in the presence
of magnetic impurities. In this case, however, the global prefactor and the
total scattering rate are altered as compared to the case of nonmagnetic
impurities since for magnetic impurities, the evaluation of the trace
$\tr[\sigma_{i}\sigma_{j}\sigma_{k}\sigma_{l}\sigma_{j}\sigma_{m}]$
allows for additional contributions including other interband scattering
processes between the electron pockets $\mathrm{e}_{1}$ and $\mathrm{e}_{2}$.

\section{Summary and Conclusions}
\label{sec:conclusions}

Recent experiments revealed the existence of $C_{4}$-magnetic phases in
hole-doped iron-based superconductors, further fueling the discussion
about the nature of the magnetic ground state of the parent compounds.
We considered a three-band model of iron-based superconductors complemented
by an incipient fourth pocket at the $M$~point and investigated
how the interplay of impurity scattering and disorder effects in a
rigid-band approach affect the magnetic ground state.

The phase diagram is governed by the interplay of nematic and planar
couplings, $g$ and $w$, respectively. If $g>\max(0,-w)$, stripe-magnetic
order with either $\vect{M}_{1}=0$ or $\vect{M}_{2}=0$ is favored,
as it has been observed in many compounds of the iron pnictide and
iron chalcogenide families. If $g<\max(0,-w)$, a double-$\vect{Q}$
state with $|\vect{M}_{1}|=|\vect{M}_{2}|$ minimizes the free energy,
and the sign of $w$ determines whether $\vect{M}_{1}\perp\vect{M}_{2}$
(spin vortex crystal, for $w>0$) or $\vect{M}_{1}=\pm\vect{M}_{2}$
(charge-spin density wave, for $w<0$) is more favorable. So far,
only the charge-spin density wave has been observed experimentally,~\cite{Allred-arxiv2015,WasserBraden-PRB2015,Mallett-EPL2015}
in contrast to theoretical models.~\cite{Chandra-PRL1990,EreminChubukov-PRB2010,Wang-PRB2015}

Although generic three-band low-energy models for the description
of FeSC allow for $C_{4}$-magnetic ground states, they leave the
spin vortex crystal (SVC) and the charge-spin density wave (CSDW)
degenerate since $w=0$. Our analysis shows that the existence of
an incipient pocket at $(\pi,\pi)$ lifts the degeneracy, however,
it would favor the formation of a spin-vortex crystal ($w>0$) and
thus cannot explain the experimental findings. The investigation of
other extensions to the three-band model such as the consideration
of additional interactions has lead to the same conclusion that the
SVC state is favorable.

Our investigation of impurity scattering, in contrast, provides a
natural explanation for the formation of a charge-spin density wave
in doped FeSC. Since the three-band model under consideration yields
$w=0$ in the absence of impurity scattering, we concentrated on the
interband scattering process between the two electron bands that can
render $w$ finite. In addition, we considered intraband scattering
in all three bands. We find $w<0$ at particle-hole symmetry as well
as for small ellipticity and detuning, suggesting that disorder can
promote charge-spin density waves. However, sufficiently large
ellipticity and detuning in combination with impurity scattering also
allow for $w>0$, i.\,e., a spin vortex crystal.

Our findings are summarized in the phase diagrams depicted in Fig.~\ref{fig:phasediagrams}
where we show the magnetic ground states that are favored in different
regimes of detuning and ellipticity. Disorder favors the double-\textbf{Q}
charge-spin density wave over the single-\textbf{Q} stripe-magnetic
SDW at small ellipticity and detuning, and increasing scattering rate
increases the parameter regime in which CSDW order is expected to
occur. 

We further investigated the effect of the dominant impurity scattering
process in FeSC, intraband scattering in the hole band, on the nematic
coupling $g$, which in the three-band model assumes a finite value
as long as the electron bands exhibit finite ellipticity. In the absence
of impurity scattering and for $\delta_{\mu}=0$, $g$ is positive,
and increasing intraband scattering in the hole band reduces the nematic
coupling constant, concordant with the experimental finding that electron
irradiation enhances the splitting between structural and magnetic
transition in the stripe-ordered phase.

Previously, controlled disorder has been proposed as a way to tune
the properties of the superconducting state in the iron-based materials.\cite{WangKreiselHirschfeldMishra-PRB2013} Analogously, our findings provide
a promising control knob to tune their magnetic ground state. In particular,
addition of impurities via electron irradiation in hole-doped compounds
near the composition where the single-\textbf{Q} to double-\textbf{Q} magnetic transition is observed could stabilize a $C_{4}$-magnetic
phase as the leading instability of the system -- currently, the $C_{4}$-magnetic phase has been mostly observed inside the $C_{2}$-magnetic
phase boundary. Similarly, removal of impurities via annealing in
samples that display the double-\textbf{Q }magnetic order could change
the nature of the $C_{4}$ phase from charge-spin density wave to
spin-vortex crystal. 

\section*{Acknowledgments}
We thank E. Berg, M. Christensen, A. Chubukov, I. Eremin, J. Kang,
S. Kivelson, M. S. Scheurer, and X. Wang for helpful discussions.
M.H. and J.S. are supported by the Deutsche Forschungsgemeinschaft
through DFG-SPP 1458 `Hochtemperatursupraleitung in Eisenpniktiden'.
A.L. acknowledges support by NSF Grant No. DMR-1606517 and in part
by DAAD grant from German Academic Exchange Services. Support for
this research at the University of Wisconsin-Madison was provided
by the Office of the Vice Chancellor for Research and Graduate Education
with funding from the Wisconsin Alumni Research Foundation. R.M.F.
is supported by the U.S. Department of Energy, Office of Science,
Basic Energy Sciences, under award number DE-SC0012336.

\end{document}